\documentclass[twocolumn]{aa} 

\usepackage{graphicx}
\usepackage{txfonts}
\usepackage{natbib} 
\bibpunct{(}{)}{;}{a}{}{,} 
\begin{document}
   \title{Direct imaging of extra-solar planets in star forming regions}

   \subtitle{Lessons learned from a false positive around IM~Lup\thanks{Based on the ESO observing programs 380.C-0910, 084.C-0444, 287.C-5040; and HST observing program~10177.}}

   \author{D.~Mawet\inst{\ref{inst1}}\fnmsep\inst{\ref{inst2}}\and O.~Absil\inst{\ref{inst3}}\fnmsep\thanks{FNRS postdoctoral researcher.}\and G.~Montagnier\inst{\ref{inst1}}\and P.~Riaud\inst{\ref{inst3}}\and J.~Surdej\inst{\ref{inst3}}\and C.~Ducourant\inst{\ref{inst5}}\and J.-C. Augereau\inst{\ref{inst4}}\and S.~R\"ottinger\inst{\ref{inst4}}\and J.~Girard\inst{\ref{inst1}}\and J.~Krist\inst{\ref{inst2}}\and K.~Stapelfeldt\inst{\ref{inst6}}}

   \institute{European Southern Observatory, Alonso de Cord\'ova 3107, Vitacura, Santiago, Chile\\
              \email{dmawet@eso.org}\label{inst1}
         \and
             	NASA-Jet Propulsion Laboratory, California Institute of Technology, 4800 Oak Grove Drive, Pasadena, CA 91109, USA\label{inst2}
	\and
		Institut d'Astrophysique et de G\'eophysique, University of Li\`ege, All\'ee du 6 Ao\^ut 17, 4000 Sart Tilman\label{inst3}
	\and
		UJF-Grenoble 1 / CNRS-INSU, Institut de Plan\'etologie et d'Astrophysique de Grenoble (IPAG), UMR 5274, Grenoble, F-38041, France\label{inst4}
	\and
		University of Bordeaux, LAB, UMR5804, 33271 Floirac Cedex, France\label{inst5}
	\and
		NASA-Goddard Space Flight Center, 8800 Greenbelt Road, Greenbelt, MD 20771, USA\label{inst6}
                         }

   \date{Received May 24, 2012; accepted July 23, 2012}

 
  \abstract
  {Most exoplanet imagers consist of ground-based adaptive optics coronagraphic cameras which are currently limited in contrast, sensitivity and astrometric precision, but advantageously observe in the near-infrared window ($1-5\, \mu$m). Because of these practical limitations, our current observational aim at detecting and characterizing planets puts heavy constraints on target selection, observing strategies, data reduction, and follow-up. Most surveys so far have thus targeted young systems ($1-100$ Myr) to catch the putative remnant thermal radiation of giant planets, which peaks in the near-infrared. They also favor systems in the solar neighborhood ($d<80$ pc), which eases angular resolution requirements but also ensures a good knowledge of the distance and proper motion, which are critical to secure the planet status, and enable subsequent characterization. }
  {Because of their youth, it is very tempting to target the nearby star forming regions, which are typically twice as far as the bulk of objects usually combed for planets by direct imaging. Probing these interesting reservoirs sets additional constraints that we review in this paper by presenting the planet search that we initiated in 2008 around the disk-bearing T~Tauri star IM~Lup, which is part of the Lupus star forming region (140-190 pc).}
  {We show and discuss why age determination, the choice of evolutionary model for both the central star and the planet, precise knowledge of the host star proper motion, relative or absolute (between different instruments) astrometric accuracy (including plate scale calibration), and patience are the key ingredients for exoplanet searches around more distant young stars.}
  {Unfortunately, most of the time, precision and perseverance are not paying off: we discovered a candidate companion around IM~Lup in 2008, which we report here to be an unbound background object. We nevertheless review in details the lessons learned from our endeavor, and additionally present the best detection limits ever calculated for IM~Lup. We also accessorily report on the successful use of innovative data reduction techniques, such as the damped-LOCI and iterative roll subtraction.}
  {}
  

   \maketitle
%

\section{Introduction}
Direct imaging constitutes an attractive technique for exoplanet detection as it provides straightforward means to characterize planets and their host system \citep{Absilmawet2010} through, e.g., orbital motion \citep{Soummer2011, Chauvin2012}, spectro-photometry of planetary atmospheres \citep{Janson2010, Galicher2011, Bonnefoy2011}, or planet-disk interactions \citep{Lagrange2012}. Direct imaging has also the potential of understanding and bridging the gap between the population of extremely close planets discovered by radial velocity or transit techniques and the free floating planets discovered by microlensing observations \citep{Sumi2011,Quanz2012}. Indeed, many exoplanet candidates directly imaged so far have projected distances up to several hundreds of AU. On the other hand, some free floating low-mass objects have been found to be kinematically associated at projected distances of thousands of AU \citep{Caballero2006}. This raises the questions of their formation and the very definition of planets, on which direct imaging is key to shed more light.

However, imaging extra-solar planets around other stars constitutes a multiple challenge, and the practical hurdles are numerous. First of all, the angular separation between planets and stars is very small (e.g. $<$500 mas for a 5-AU distance at 10 pc), usually requiring diffraction limited capabilities on 8-meter class telescopes \footnote{Note the exception presented in \citet{Serabyn2010}, who showed a snapshot of 3 out of the 4 planets of HR~8799 taken with an adaptively-corrected 1.5-meter telescope and a next-generation vector vortex phase mask coronagraph \citep{Mawet2010}.}. Second, the contrast between a planet and its host star ranges from $\simeq 10^{-3}$ for hot giant planets in the infrared to $\simeq 10^{-10}$ for Earth-like planets in the visible. The contrast issue requires exquisite image (hence wavefront) quality to feed coronagraphic devices, most of the time very specialized observing strategies \citep[e.g., angular differential imaging or ADI,][]{Marois2006}, and corresponding data reduction techniques such as the Locally Optimized Combination of Images \citep[LOCI,][]{Lafreniere2007}.

Once a faint point source has been detected around a star, pointing to the potential discovery of a companion candidate, precise differential astrometric monitoring of the latter needs to be carried over a sufficiently long time so that the stellar proper motion overcomes the astrometric precision of the detected object by a sufficient margin (if the object is bound, it moves with its host star).  \citet{Neuhauser2012} also argues that a spectrum, when possible (proximity to the host star often prevents to take clean uncontaminated spectra), can determine the spectral type and temperature of the companion, and thus indicates a planetary mass or sub-stellar body, but still possibly a cool background object. Both tests might sometimes be necessary, especially when targeting young associations where objects can potentially share common proper motion, likely to be small and rather uncertain (at the distance of star forming regions), making this astrometric process more difficult and the required time baseline longer. The T~Tauri star ScoPMS~214 is a typical example, where a candidate companion was shown to share common proper motion, but was spectroscopically identified as a foreground M dwarf \citep{Metchev2009}. In young associations, the probability for small and/or shared proper motion is thus significant. A third possibility for confirming the bound character of the companion is the detection of the orbital motion, but that implies that the candidate is on a reasonably tight orbit (period $<$ 1000 years), in order to be sampled with sufficient accuracy over a time baseline of a few years.

Most of the objects imaged so far are orbiting young stars (see exoplanet.eu for a thorough and up-to-date list, and \citet{Neuhauser2012} for a recent detailed review). Youth is the current bias of high contrast imaging, as short period, inclination or distances (orbital and/or parallactic) are the biases of radial velocity, transit and micro-lensing techniques, respectively. Indeed, the thermal radiation of young exoplanets peaks in the near-infrared, making them more easily detectable by several orders of magnitude than if we were to observe them in reflected light in the visible. Since the detected emission comes from the intrinsic thermal radiation of the planet, its physical properties (temperature, mass and radius) can only be inferred based on cooling track models, which critically depend on age and formation mechanisms/history \citep{Allard2003,Marley2007,Fortney2008, Spiegel2012}. Due to this very high sensitivity to initial conditions, deriving the nature of low-mass objects and young planets is indeed more problematic than for older ones, especially for long-period companions where the dynamical mass is more difficult to infer. For instance, \citet{Stassun2006} found a young eclipsing brown-dwarf binary in which the cooler object is the more massive one, which is very surprising, because most theoretical models predict that a brown dwarf of a given mass will at all times be warmer than a lower-mass brown dwarf of the same age. \citet{Stassun2008} also found a binary where both stars have the same mass within 2\% but their surface temperatures differ by 300K. Therefore, dynamical mass, hence astrometric precision and proper motion knowledge, age, together with distance, become key parameters that will determine the final precision and confidence on the companion physical characteristics.

In the present paper, we discuss these issues in details, and illustrate them with our search for planetary candidates around the young T~Tauri star IM~Lup. The paper is organized as follows: after presenting the IM~Lup system in Sect.~\ref{sec:imlup}, we describe our VLT/NACO 2008 observations and discovery of a putative candidate companion in Sect.~\ref{sect:obs}. Then, in Sect.~\ref{sec:preimg}, we present our re-analysis of the pre-discovery data taken with HST/NICMOS in 2005, followed by epoch 3 and 4 images taken with VLT/NACO  in 2010 and 2011 in Sect.~\ref{sec:epoch34}. In Sect.~\ref{sec:discussion}, we discuss our discovery in details and underline the difficulties of planet searches around this type of objects, in terms of astrometry and related proper motion analysis. We also establish benchmark detection limits around IM~Lup after reestablishing its age (Sect.~\ref{subsect:age}), before concluding in Sect.~\ref{sec:conclusion}.
 \begin{table}
      \caption[]{Fundamental properties of IM~Lup and associated optically thick circumstellar disk.}
         \label{table0}
	$$
         \begin{array}{p{0.5\linewidth}l}
            \hline
            \noalign{\smallskip}
            Properties      &Value \\
            \noalign{\smallskip}
            \hline
            \noalign{\smallskip}
          Names			& {\rm IM~Lup, Sz82, PDS 75} \\
          				& {\rm IRAS 15528-3747} \\
	Spectral type		& {\rm M0}			\\
	Temperature		& {\rm 3900 K} 		\\
	Class			& {\rm CTTS / WTTS}	\\
	Age				& 0.5-1.75 {\rm \, Myr}^{\mathrm{a}}	\\
	Association		& {\rm Lupus}		\\
	Distance			& 140-190 {\rm \, pc}^{\mathrm{a}}	\\
	V mag			& 11.96		\\
	H mag			& 8.089		\\
	K mag			& 7.739		\\
	Lp mag			& 7.29		\\ 
	Disk radius$^{\mathrm{b}}$	& 0.32-400\, {\rm AU}	\\
	Disk  inclination$^{\mathrm{b}}$ 	& \simeq 50^\circ \\ 
	Disk-to-star mass ratio$^{\mathrm{b}}$		&\simeq 0.1	    \\
	Grain sizes$^{\mathrm{b}}$		&0.03-3000\, \mu m	    \\
            \noalign{\smallskip}
            \hline
         \end{array}
	$$
\begin{list}{}{}
\item[$^{\mathrm{a}}$] See the discussion in Sect.~\ref{subsect:age}.
\item[$^{\mathrm{b}}$] Taken from the best fitted model presented in \citet{Pinte2008}.
\item[-] References for all other values are in the text.
\end{list}
   \end{table}


\section{IM~Lup: a young T Tauri star with a massive circumstellar disk}\label{sec:imlup}

IM~Lup (Table~\ref{table0}) is a young M0 (T$=3900$ K) T~Tauri star (TTS) with an equivalent width of the H$\alpha$ emission known to vary from 7.5 to 21.5$\AA$, confirming its status as a borderline weak-line/classical TTS. Part of the Lupus association (140-190 pc, see Sect.~\ref{subsect:age}), it is one of four young stellar objects in the small $^{13}$CO$(1-0)$ Lupus 2 core near the extreme T Tauri star RU Lup \citep{Tachihara1996}. Our age estimate described in Sect.~\ref{subsect:age} yields 0.5-1.75 Myr.

\begin{table*}
\caption{Observing log for IM~Lup and the reference stars used in this work.}             
\label{table1}      
\centering          
\begin{tabular}{lccccccccc}     
\hline\hline       
Target &Prog.ID & $\alpha$ (J2000)& $\delta$ (J2000) &Filter & UT date &Exp.~time &Tel./instr. &Strat.$^{\mathrm{a}}$ & Strehl (\%)$^{\mathrm{b}}$   \\ 
\hline
IM~Lup &10177 &15h 56' 09" &-37$^{\circ}$ 56' 06" &F160W &29/03/05  &$1350s$ &HST/NICMOS &ADI &-   \\
IM~Lup &380.C-0910(A) &15h 56' 09" &-37$^{\circ}$ 56' 06" &Ks &29/03/08 &$1350s$ &VLT/NACO &RDI & $58\pm 5$     \\
CD-37 8989 &380.C-0910(A) &13h 54' 27" &-38$^\circ$ 14' 54" &Ks &29/03/08 &$1200s$ &VLT/NACO &- & $50\pm 3$    \\
CD-35 9033 &380.C-0910(A) &13h 50' 00" &-36$^\circ$ 33' 40" &Ks &30/03/08 &$1200s$ &VLT/NACO &- & $48\pm 3$    \\
IM~Lup &084.C-0444D &15h 56' 09" &-37$^{\circ}$ 56' 06" &Lp &19/04/10 &$1350s$  &VLT/NACO &RDI &$80\pm 5$     \\
LHS3286 &084.C-0444D  &17h 23' 49" & -32$^{\circ}$15' 16" &Lp &19/04/10 &$1350s$  &VLT/NACO &- &$85\pm 5$    \\
IM~Lup &287.C-5040(A) &15h 56' 09" &-37$^{\circ}$ 56' 06"  &Ks &22/07/11 &$1350s$ &VLT/NACO &I &$40\pm 3$     \\
IM~Lup &287.C-5040(A) &15h 56' 09" &-37$^{\circ}$ 56' 06"  &H &25/07/11 &$1350s$ &VLT/NACO &I &$30\pm 3$     \\
\hline                  
\end{tabular}
\begin{list}{}{}
\item[$^{\mathrm{a}}$] Observing strategy: ADI (angular differential imaging), RDI (reference star differential imaging), I (imaging).
\item[$^{\mathrm{b}}$] The Strehl ratio was measured on the reduced image, or on the acquisition PSF (for saturated or coronagraphic data). Seeing and coherence time ($\tau_0$) conditions in the Visible ($\simeq 0.5$ $\mu$m) were very good for all observations, typically $0\farcs 6-0\farcs 8$, and $3-8$ ms, respectively.
\item[-] Note that all targets were observed at airmass $\simeq 1.1$, except for HST of course.
\end{list}
\end{table*}

Despite the low accretion-related activity of IM~Lup \citep{Reipurth1996, Wichmann1999}, long wavelength observations from the millimeter \citep{Nuernberger1997,vanKempen2007,Lommen2007} to the infrared \citep{Padgett2006} reveal ample evidence for gas-rich circumstellar material in the system. IM~Lup's protoplanetary disk scattered light was imaged in 1999 in the visible with HST/WFPC2 (PI: Stapelfeldt, Prog.~ID~7387). It was followed in the near-infrared by HST/NICMOS images obtained in 2005 (PI: G.~Schneider, Prog.~ID~10177). An extensive modeling study of the IM~Lup disk was performed by \citet{Pinte2008}, using multi-wavelength spectro-photometry and images in a global fit with a 3D radiative transfer model, and led to quantitative evidence for dust processing and evolution in the disk.

A more recent paper by \citet{Panic2009} presents SMA (Submillimeter Array) observations showing a break in the gas and dust surface density of the IM~Lup disk, seen to extend much further than the 400 AU outer edge determined by \citet{Pinte2008}. One of the proposed explanations for the break is a companion body near the break at 400 AU. Indeed, a companion of 1 $M_{\rm Jup}$  could open a gap in the disk and affect its spreading. \citet{Panic2009} however argue that no candidate companion at this separation is visible in the HST image of \citet{Pinte2008}.

\section{Discovery of a candidate companion with VLT/NACO in 2008}\label{sect:obs}
As part of a coronagraphic study of young stars (prog.~ID 380.C-0910(A), PI: Mawet), we observed IM~Lup in March 2008 with NAOS-CONICA, the adaptive optics (Nasmyth Adaptive Optics System) and near-infrared spectrograph and imager of the Very Large Telescope (VLT).

\subsection{Observing strategy for the 2008 discovery data set}
For our discovery image in 2008, we used the four-quadrant phase-mask (FQPM) coronagraph \citep{Rouan2007} in the Ks band. The FQPM is a phase-mask coronagraph applying a $\pi$ phase shift between adjacent quadrants. The starlight, when centered on the FQPM cross-hair, undergoes a destructive interference upon propagation to the downstream pupil plane.

All of our frames for 2008 were taken with the Ks filter and the S13 camera (13.27 mas/pixel). This fine sampling (4 pixels per resolution element $\lambda/d$, where $\lambda$ is the observing wavelength and $d$ the telescope diameter) was necessary to center the target star on the FQPM cross-hair precisely. The main calibrator stars were carefully selected to present roughly the same V-K color as the target (Table \ref{table1}). Matching the V magnitudes is important to ensure similar AO corrections between the target and reference stars, as the visible wavefront sensor of NAOS is mostly sensitive at V and R. Ks magnitudes have also to be matched to ensure SNR matching for the quasi-static speckles. Also, to avoid flexure-induced semi-static speckle variations as much as possible and to ensure a consistent telescope orientation with respect to the instrument between the target and the reference, the calibration stars were chosen and observed at the same parallactic angle as the target star. This condition was met on a best effort basis since the availability of a suitable reference fulfilling the set of constraints is never guaranteed, which is one of the drawbacks of the reference star differential imaging strategy (RDI).

The observing conditions for IM~Lup and the reference stars were very good with a visible seeing between $0\farcs 6$ and $0\farcs 8$. The total integration time was about 1350 sec and 1200 sec for each target (see Table \ref{table1}). Respecting the consideration discussed here above to calibrate time-dependent PSF variations (speckle), we acquired coronagraphic images of the reference star 90 min after the science target at roughly the same parallactic angle. To reduce drift and pupil rotation, the centering was checked and corrected every 80 sec.
\begin{figure*}[!ht]
  \centering
\includegraphics[scale=0.5]{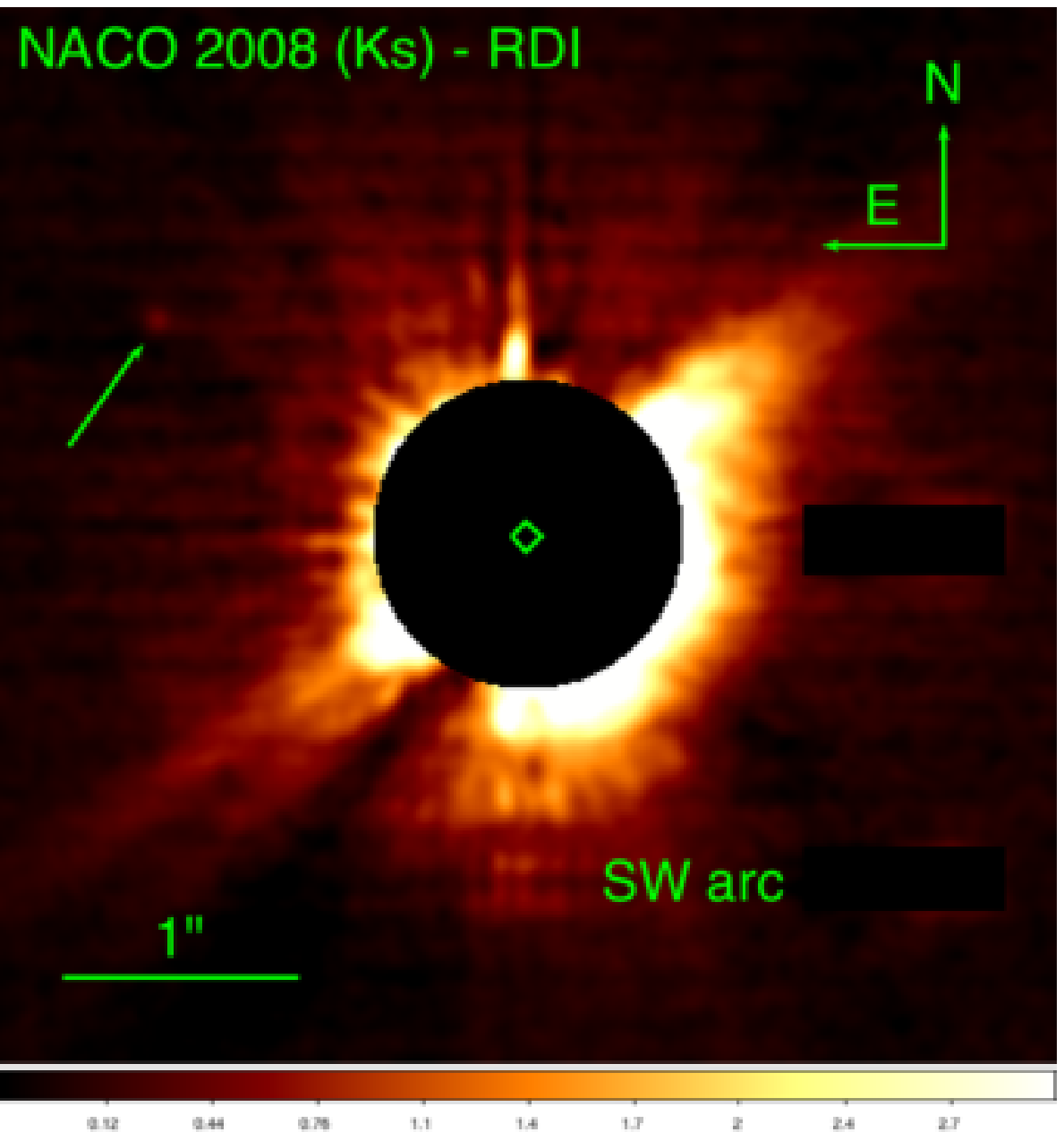}
\includegraphics[scale=0.5]{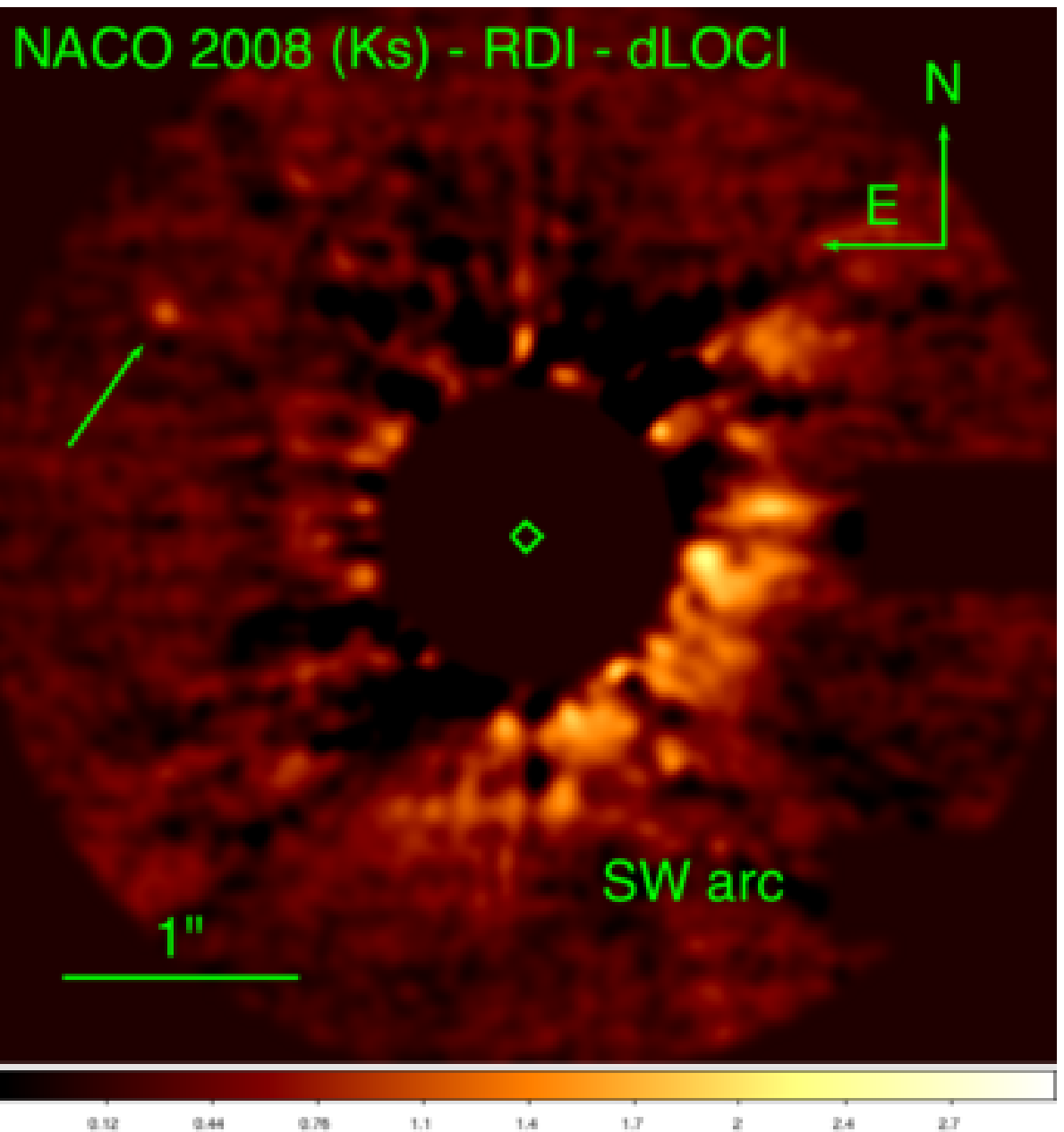}
\includegraphics[scale=0.5]{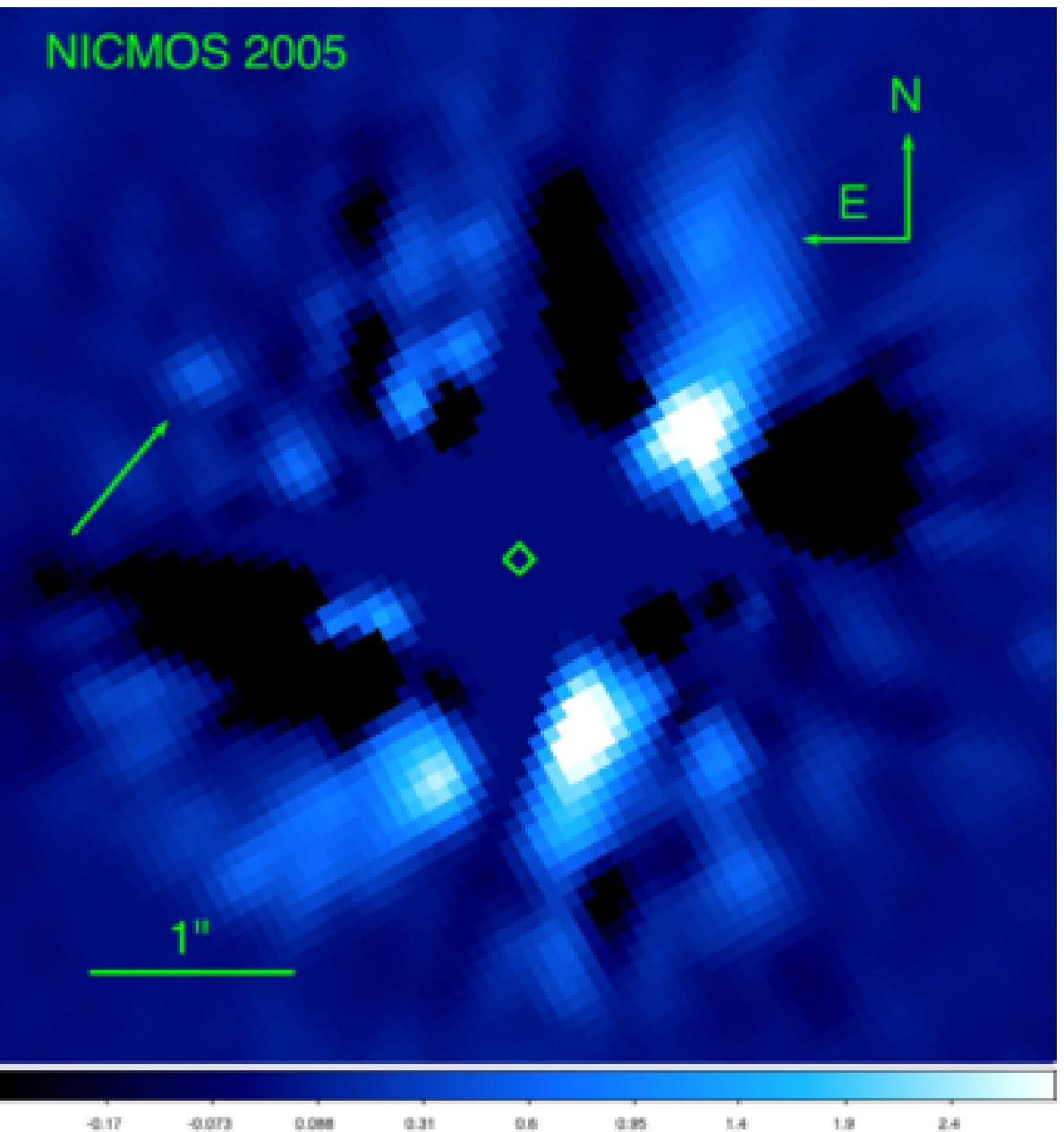}
\includegraphics[scale=0.5]{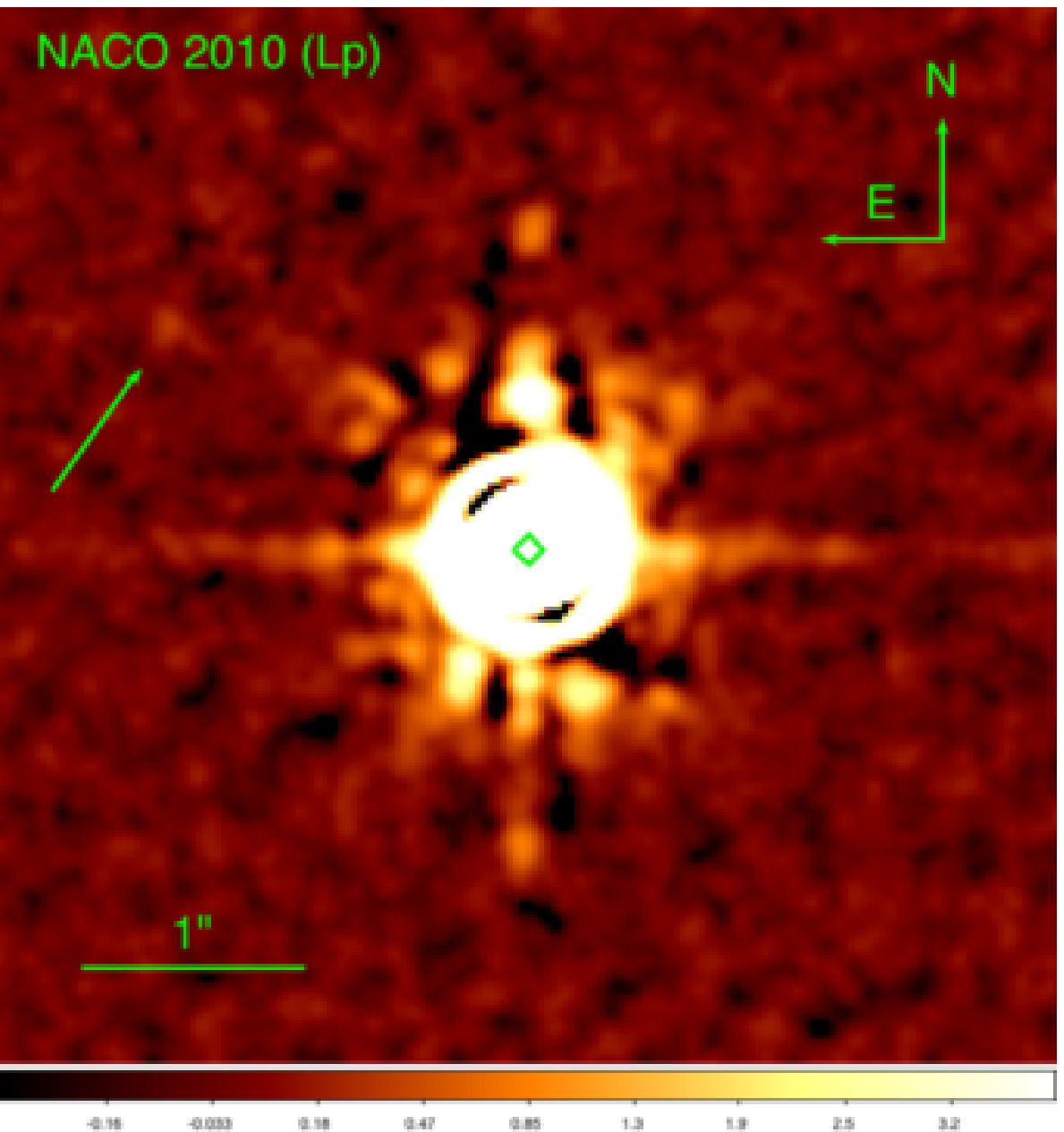}
\includegraphics[scale=0.5]{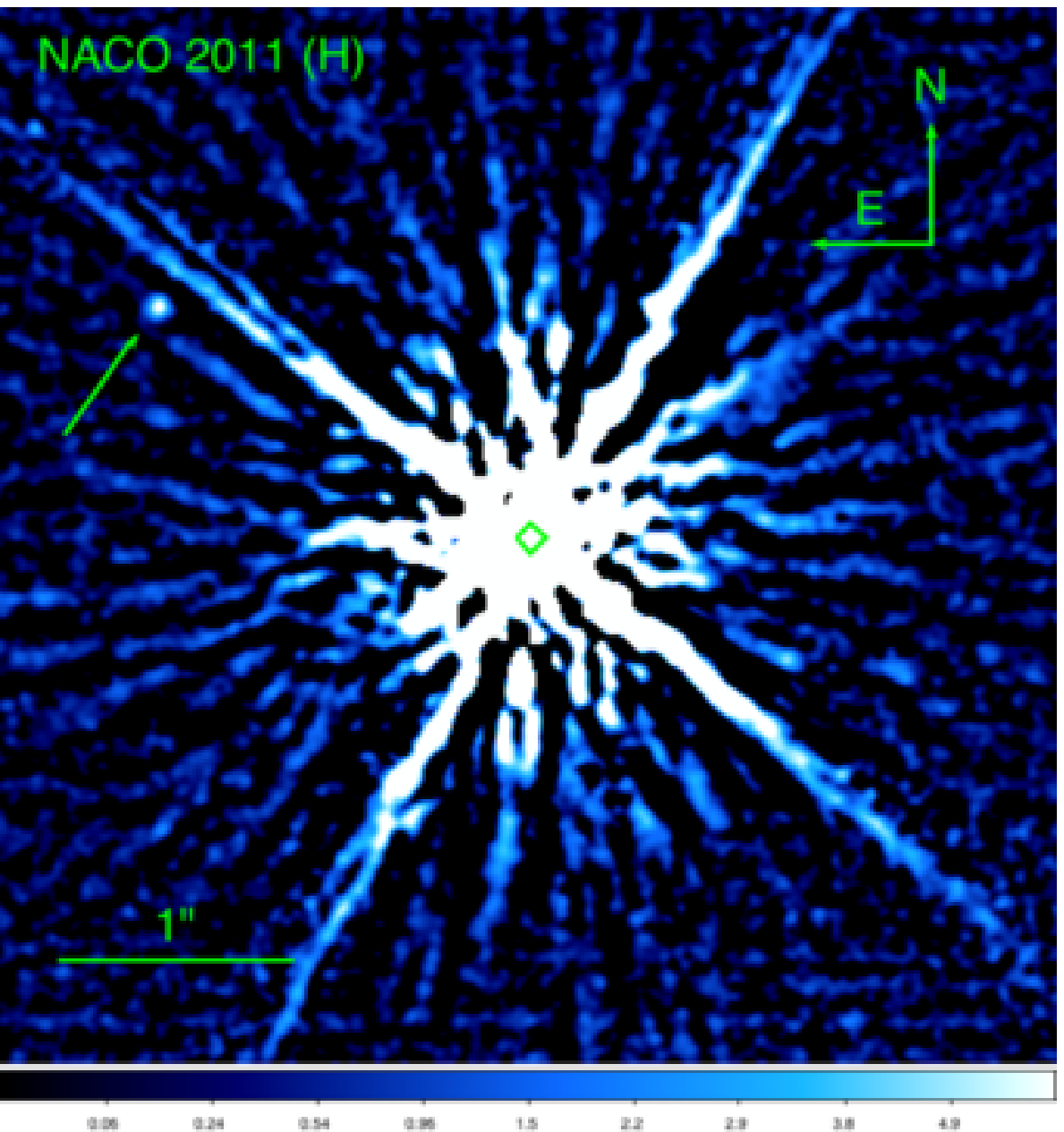}
\includegraphics[scale=0.5]{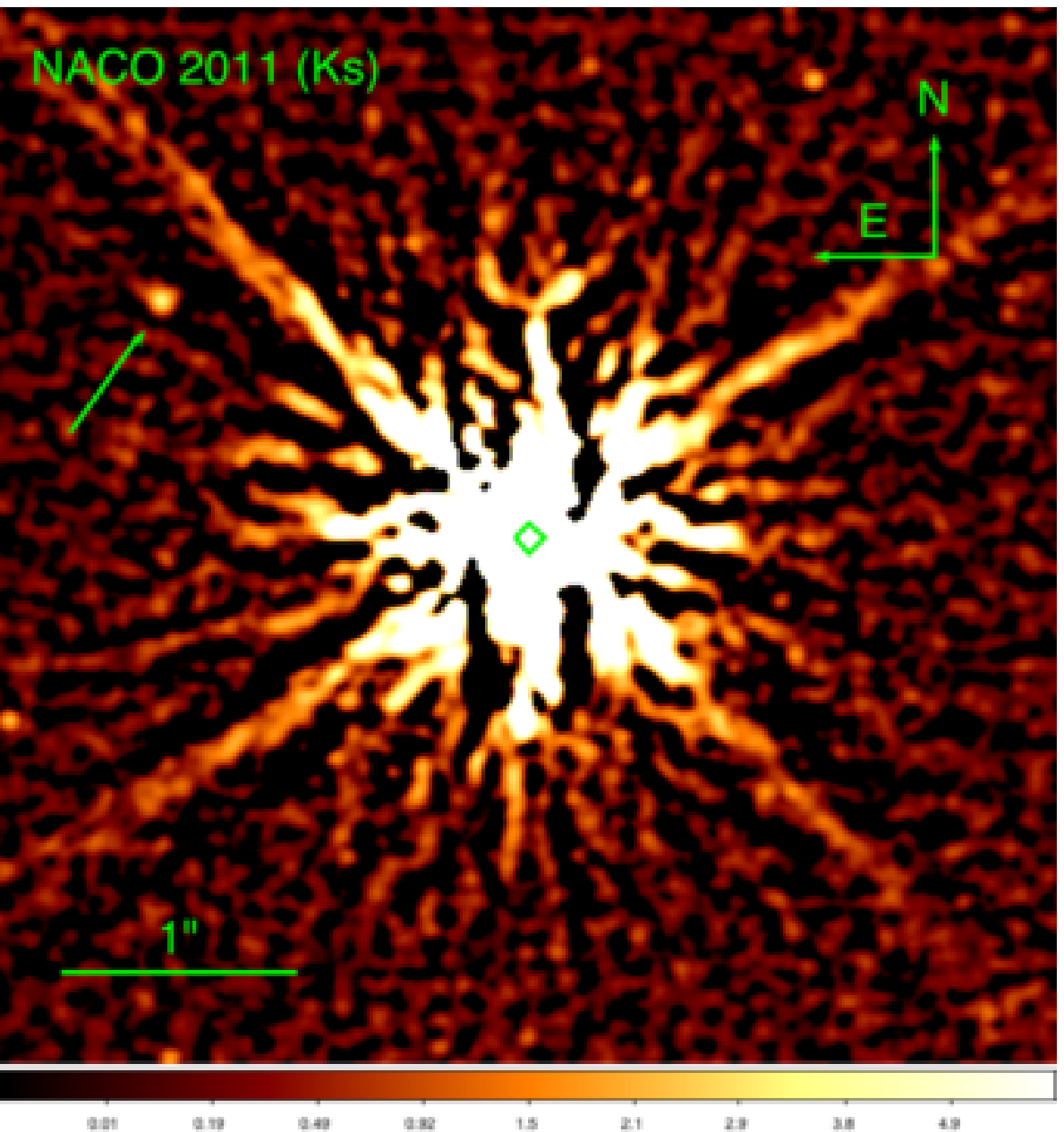}
  \caption{The upper strip presents the simple reference star subtracted Ks-band image (left) and the damped-LOCI processed image (middle) obtained with NACO in 2008, which show a very faint off-axis point source ($m_{Ks}\simeq 19.1$) to the northeast of IM~Lup. The upper right image presents the 2005 F160W NICMOS observations, using an iterative roll-subtraction technique. The lower strip presents the NACO Lp-band image of 2010 (left), the 2011 NACO H-band image (middle), and the 2011 NACO Ks-band image (right). The off-axis point source is detected in all images (identified with an arrow). \label{sz82_images}}
\end{figure*}

\subsection{Data analysis for the 2008 discovery data set}\label{subsect:dataanalysis}
The data reduction proceeds as follows. NACO coronagraphic acquisition template moves the telescope alternatively between a fixed object position and a jittered set of sky positions which are median combined and subtracted to the object, removing the background and dark contributions at the same time. The normalization of the resulting image with the flat provides the first stage of data processing. However, this treatment does not completely remove the electronic noise of the readout process nor the two electronic ghosts which are characteristic of CONICA.

The second stage of the data reduction process consists in co-adding of the images with a sub-pixel centering procedure. For that, we applied a hybrid method which correlates the centroid of the unsaturated coronagraphic pattern with a pre-computed look-up table based on a diffraction model of the FQPM. Using this sophisticated method, we routinely achieve a centering precision of $\sigma=0.1$ pixel or 1.4 mas rms. Despite the coronagraph starlight 10-fold attenuation, scattered starlight still dominates the extended source flux. Since the primary objective of our original program was disk imaging, and given the relative novelty of the ADI technique at the time of the first observations, and the fact that ADI is not an optimal strategy for nearly face-on disk (Milli et al.~2012, accepted to A\&A), we chose to use classical RDI.

The final VLT/NACO 2008 image (see Fig.~\ref{sz82_images}, top middle) was obtained using an enhanced version of the LOCI  algorithm \citep{Lafreniere2007}. In short, LOCI finds the optimal linear combination of reference frames (here from the two reference stars CD-37~8989 and CD-35~9033) to minimize the noise in a given zone of the target image. The process is repeated until the area of interest in the target image is completely reduced. LOCI in its original form was conceived to find point sources, and has a known tendency to attenuate signal from extended sources such as circumstellar disks. However, this defect of the generic LOCI algorithm can be brought under control by a fine tuning of the geometrical parameters such as the size of the optimization zone, the number of reference frames used in the correlation matrix, as well as the introduction of a damping parameter (Lagrange multiplier) to balance flux conservation with noise attenuation as in \citet{Pueyo2012}.

\begin{table*}
\caption{Astrometry of the point source to the NE, along with its relative and absolute photometry.}
\label{table2}
\centering          
\begin{tabular}{lcccccc}
\hline\hline
Data set 			&Filter 	&SNR 		&$\Delta \alpha$ ($\arcsec$)	 &$\Delta \delta$ ($\arcsec$)		&m 					&M$^{\mathrm{a}}$ \\ \hline
NICMOS 2005 		&F160W	&$\simeq 3$     	&$1\farcs 526\pm 0\farcs 043$	 &$0\farcs 937 \pm 0\farcs 043$	&$19.3^{+0.4}_{-0.3}$ 	&$13.2^{+0.8}_{-0.7}$ \\ 
NACO 2008		&Ks		&$\simeq 10$   	&$1\farcs 541\pm 0\farcs 007$	 &$0\farcs 936 \pm 0\farcs 007$	&$19.1^{+0.2}_{-0.25}$ 	&$13.0^{+0.4}_{-0.5}$ \\ 
NACO 2010		&Lp		&$\simeq 2$      &$1\farcs 570\pm 0\farcs 020$  &$0\farcs 978 \pm 0\farcs 020$	&$17.8^{+0.7}_{-0.6}$	&$11.7^{+1.4}_{-1.2}$ \\ 
NACO 2011		&Ks		&$\simeq 10$   	&$1\farcs 575\pm 0\farcs 009$	 &$1\farcs 002 \pm 0\farcs 009$	&$19.2^{+0.2}_{-0.35}$	&$13.1^{+0.4}_{-0.7}$ \\ 
NACO 2011$^{\mathrm{b}}$	&H		&$\simeq 5$   	&$1\farcs 594\pm 0\farcs 025$	 &$0\farcs 996 \pm 0\farcs 025$	&$19.25^{+0.3}_{-0.2}$ 	&$13.2^{+0.6}_{-0.4}$\\ \hline
\end{tabular}
\begin{list}{}{}
\item[$^{\mathrm{a}}$] Assuming physical association. The error bars include the distance uncertainty (140-190 pc).
\item[$^{\mathrm{b}}$] The error bars are large due to the presence of the diffraction pattern of the telescope spiders close to the point source.
\end{list}
\end{table*}

\subsection{Candidate companion and new image of the disk}
In our 2008 Ks-band dataset (Table \ref{table1}), we detected a very faint companion at a signal-to-noise ratio (SNR) of $\simeq 10$ (Fig.~\ref{sz82_images}, top left). Applying the damped-LOCI of \citet{Pueyo2012}, both the companion and the disk SW arc are detected simultaneously with minimum contamination from starlight scattering (Fig.~\ref{sz82_images}, top middle). For the first time, we note that damped-LOCI, originally invented to detect point-source in multi-spectral data, can be successfully applied to the detection of circumstellar disks, improving upon the original LOCI of \citet{Lafreniere2007}.

We performed relative astrometry on the final reduced image, using centroiding and a specific pre-computed look-up table for the star position behind the FQPM (see Sect.~\ref{subsect:dataanalysis}), and gaussian fitting for the candidate companion (with a subpixel precision). The companion is located to the North-East of IM~Lup, at a radius of 
$\simeq 1\farcs 8$, and a position angle (PA) of $\simeq 58^\circ$ (see Table~\ref{table2}).

We then performed aperture photometry using the function APER of IDL, and found a relative Ks magnitude of $19.1^{+0.2}_{-0.25}$. Naively assuming that the point source is physically associated to IM~Lup (Sect.~\ref{sec:discussion}), this corresponds to an absolute Ks magnitude of $13.0^{+0.4}_{-0.5}$, where the uncertainty is mostly due to the poor knowledge of the star distance (140-190 pc, see Sect.~\ref{subsect:age}).

Note that the disk is detected as an arc to the SW, extending up to $\simeq 1\farcs 8$ along the major axis, and $\simeq 1\farcs 3$ along the minor axis, consistent with previous HST observations. We also measure a position angle of $140^\circ \pm 10^\circ$ for the major axis, which is consistent with the value reported in \citet{Pinte2008}.

\section{Pre-imaging of the candidate companion with HST/NICMOS in 2005}\label{sec:preimg}
Following the NACO discovery, we retrieved the archival HST/NICMOS data of IM~Lup obtained in 2005 in the F160W filter (PI: G.~Schneider, Prog.~ID~10177). The reason for the non-detection of the companion by successive groups and notably by \citet{Pinte2008} is rather simple: they primarily aimed at characterizing the extended structure, and applied corresponding soft data reduction techniques (single reference star subtraction). 

Taking advantage of the availability of two images taken at 2 different roll angles separated by 30 degrees, we used the iterative roll-subtraction algorithm introduced in \citet{Heap2000}, and described in \citet{Krist2010}. A simple roll-subtracted image contains a positive image of a companion and a negative one rotated by an angle equal to the difference between the telescope orientations. Ideally, these two would be combined to form a single, positive image. Instead of directly subtracting images from the two orientations and then trying to combine the results, an iterative technique can be used that solves for those portions of the two unsubtracted images that appear static on the detector (i.e., the PSF) and those that appear to rotate as the telescope rolls (i.e., the sky, including any companion or disk).

This method, which has been successfully used for $\beta$ Pictoris \citep{Heap2000} and HD~207129 \citep{Krist2010}, clearly ($\simeq 3\sigma$) reveals the companion in the NICMOS data as well (Fig.~\ref{sz82_images}, top right). The position of the star behind the opaque Lyot coronagraph of NICMOS was determined by two different methods. The first position was obtained by iteratively shifting a Tiny Tim model \citep{Krist1993}
PSF by subpixel amounts until the residuals were visibly minimized. The second position estimate was obtained by measuring the intersection of the diffraction spikes of the secondary mirror support structures. Both methods yield similar result within half a pixel, which we considered as a systematic error folded in the error bar calculation (Table \ref{table2}).

\section{Epoch 3 (2010) and 4 (2011) with VLT/NACO}\label{sec:epoch34}
Given the insufficient time baseline and astrometric precision of the NICMOS data point, we waited for a few years and re-observed IM~Lup in 2010 (Lp band) and in 2011 (H and Ks band), using NACO again (Table \ref{table1}). The Lp-band data only marginally (SNR$\simeq 2$) shows the companion (Fig.~\ref{sz82_images}, bottom left), with an estimated Lp magnitude of $17.8^{+0.7}_{-0.5}$. On the other hand, it is easily detected in H and Ks bands in the 2011 data (Fig.~\ref{sz82_images}, bottom middle and right). The strategy we chose for the most recent data set was to perform simple saturated-unsaturated imaging in order to enable precise astrometric and photometric analysis.

Results of the astrometry of the candidate companion relative to the host star are presented in Table~\ref{table2}. Note that the relative astrometry is somewhat different between both filters. Slight differences are expected, due for instance to differential aberrations between filters, and the difference in data quality (the Strehl ratio is naturally lower in the H band). However, it appears that the H-band astrometry is affected by the presence of the diffraction pattern of the telescope spiders (secondary mirror support structure), close to the point source. For these reasons, we only retained the Ks band astrometry in our final proper motion analysis of Sect.~\ref{sec:astrometry}, which also corresponds to the filter of the plate scale calibration described below.

\section{Discussion}\label{sec:discussion}
In this section, we elaborate on the difficulty of exoplanet candidate confirmation and characterization for young distant stars.

\begin{table}
\caption{Proper motions for IM~Lup found in major astrometric catalogues}
\label{table3}
\centering
\begin{tabular}{lcccccccc}
\hline\hline 
Origin$^{\mathrm{a}}$ 		& $\mu_{\alpha}$$^{\mathrm{b}}$  		& $\mu_{\delta}$$^{\mathrm{b}}$ 	&$|\mu|$$^{\mathrm{c}}$ \\
          		& (mas/yr)         		& (mas/yr)      		 & (mas/yr)\\  
\noalign{\smallskip}\hline \noalign{\smallskip}      
Hipparcos 	&-56.66$\pm$15.41		&-49.97$\pm$7.31	&76$\pm$17\\
Leeuwen 		&-35.5$\pm$21.81		&-22.93$\pm$14.98	&42$\pm$26\\
PMS$^{\mathrm{*}}$ 		&-3$\pm$2			&-21$\pm$2		&21$\pm$3\\
SPM4 		&-12.7$\pm$3.9		&-21.5$\pm$4		&25$\pm$6\\
PPMXL 		&-2.03$\pm$4			&-21.53$\pm$4		&22$\pm$6\\
UCAC3 		&-15.4$\pm$4.7		&-22.6$\pm$5.1	&27$\pm$7\\
\hline
\end{tabular}
\begin{list}{}{}
\item[$^{\mathrm{a}}$] Column (1) gives the origin of the measurement: \textit{Leeuwen}: re-reduction of Hipparcos data \citep{Leeuwen2007}, \textit{PMS} \citep{Ducourant2005}, \textit{SPM4}: the Yale/San Juan SPM4 Catalog (2009 http://www.astro.yale.edu/astrom/spm4cat/spm4.html), \textit{PPMXL} \citep{Roeser2010}, UCAC3 \citep{Zacharias2010}.
\item[$^{\mathrm{b}}$] Column (2) and (3) give proper motions together with uncertainties.
\item[$^{\mathrm{c}}$] Column (4) gives the modulus of proper motion.
\item[$^{\mathrm{*}}$] Adopted in this work.
\end{list}
\end{table}

\subsection{Astrometry and proper motion}\label{sec:astrometry}
To test the hypothesis of an object linked to IM~Lup we can use astrometry. If the companion is linked to IM~Lup, no significant variation of their angular separation should be observed along time due to proper motion, since they will be co-moving. The only cause of variation would result from the orbital motion of the companion around the center of mass of the system. The apparent angular separation of $\simeq 1\farcs 8$ corresponds to a deprojected physical separation of about 350-480 AU at 140-190 pc. We can expect a planet orbiting at such a distance from its central star to have extremely long periods (several thousand years at the minimum) so that its orbital motion is not detectable in our astrometric observations spread over 7 years.

\subsubsection{Prior astrometric calibration of CONICA}\label{subsec:astroconica}
To calibrate the NACO plate scale and detector orientation in a consistent and precise way between the 2008 and 2011 epochs, we used the star clusters Theta Orionis/``Trapezium'' (2008) and Omega Centauri  (2011). The reference positions of the stars in the each cluster were derived in a different program (Montagnier et al. 2012, in preparation). For each cluster, many images were taken at various positions and orientations to establish a distortion solution of the NACO plate-scale (see the method described by \citet{Anderson2003}). The linear terms of the distortion (detector axes orientations and pixel dimensions) were then derived by observing appulses of transneptunian objects. Finally, the star positions were derived with an accuracy better than 1 mas (about 200 object for the Omega Centauri cluster field, and about 50 for the Theta Orionis cluster). On the calibration images of the 2 epochs needed in our astrometric analysis, the position of the centroid $(s_x,s_y)$ of each non-saturated star (about 20 stars in the Theta Orionis's 2008 epoch, and about 40 for the Omega Centauri's 2011 epoch) on the reduced image of the field is measured in pixels; these values are then compared to the position on the sky $(\rho_x, \rho_y)$ in arcseconds with the following equations: \begin{eqnarray} \rho_x &=& x_0 + p_x s_x \cos \theta_x
- p_y s_y \sin \theta_y
\\ \rho_y &=& y_0  + p_x s_x \sin \theta_x + p_y s_y \cos \theta_y
 \end{eqnarray} where $p_x$ ($p_y$) is the plate scale along the $x$- ($y$-) axis, $\theta_x$ ($\theta_y$)
are the orientations of the detector on sky along the $x$- ($y$-) axis, and $x_0$ and $y_0$ are offsets giving the correspondence between the absolute positions (it is only used to solve the equation). A Levenberg-Marquardt minimization is then applied to find the plate scale solution. Using this calibration method, the final precision on the plate scale is $\simeq 50 \mu$as, and $\pm 0.05^\circ$ on the detector orientation. Note that this accuracy is only one item in the error budget of the final astrometric precision, which depends on many other terms all summed quadratically, such as the precision of the star position determination, fit of the off-axis location (itself dependent on the SNR), etc.

\subsubsection{Proper motion of IM~Lup}
We have considered the proper motions of IM~Lup available in the literature (see Table~\ref{table3}) to test our hypothesis of co-moving objects. We notice that the measured proper motions vary in a large range from one author to the other. The putative binary nature of IM~Lup wrongly reported by Hipparcos \citep{Wichmann1998,Lasker1996,Kharchenko2009} indicates a potential disturbance in the measurements that led to a poor astrometry. Indeed, our observations do not reveal the second component down to a magnitude $M_{Ks}\simeq 19$, indicating that the Hipparcos detection, and other reports of the binarity might have been potentially contaminated by the presence of the optically thick circumstellar disk of IM~Lup, as already suggested in \citet{Pinte2008}.

Other authors have measured the proper motion of this unresolved object with unequal precisions, leading to a consensual value in declination ($\sim$22.4 mas/yr) while the proper motion in right ascension varies largely between authors (from -3 to -15.5 mas/yr for values with reasonable precisions). The origin of such discrepancies is difficult to pinpoint since proper motion quality is not only related to the time base but also to the number of different epochs of observation and evidently to the quality of each epoch measurement. In the case of pre-main sequence stars the situation is even more difficult since depending on the target, the object may be embedded into a dust and/or gas cloud (which is the case for IM~Lup) perturbing the photo-center measurement. Moreover the morphology of the cloud may vary with time and lead to variable photo-centers at different epochs. In the data presented in Table~\ref{table3}, we chose to adopt the third one (PMS), but the four last values may be considered (PMS, SPM4, PPMXL and UCAC3) for the astrometric test of co-moving objects.

\subsubsection{Bound or not bound?}
\begin{figure*}
  \centering
\includegraphics[scale=0.6]{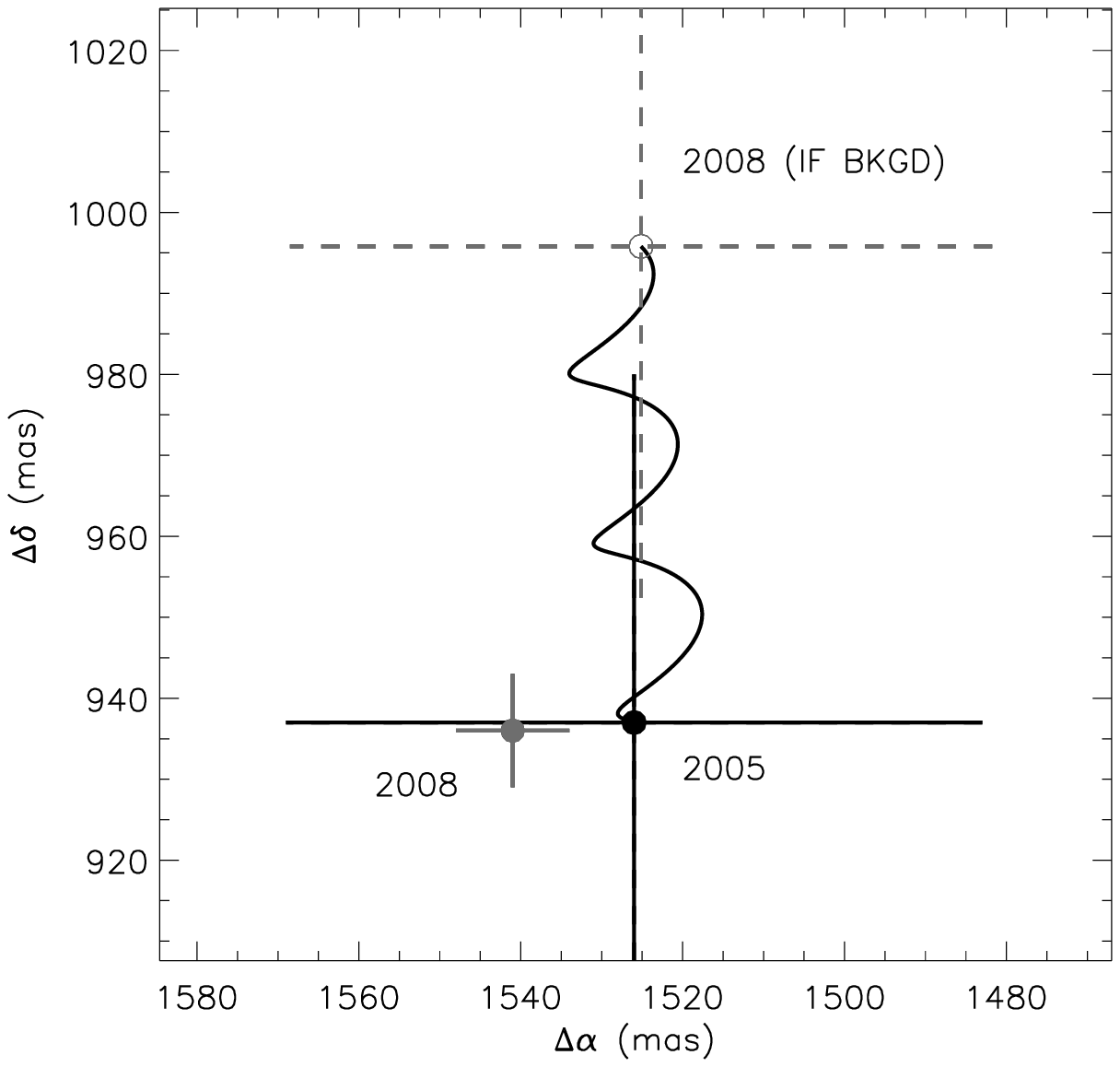}
\includegraphics[scale=0.6]{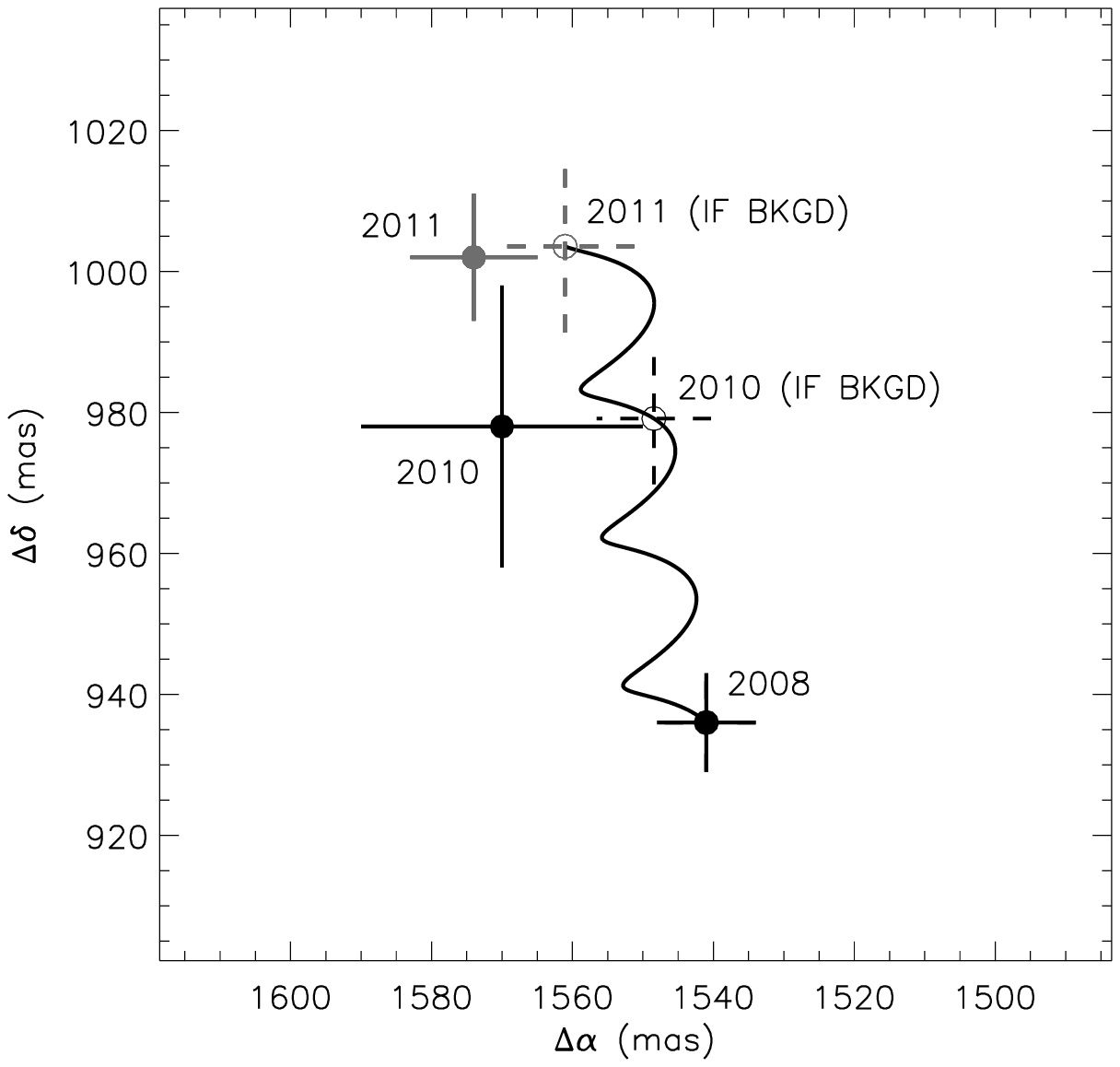}
  \caption{Left: astrometric analysis of the candidate companion proper motion between 2005 and 2008, using NICMOS and NACO. Right: astrometric analysis of the candidate companion proper motion between 2008 and 2011 (and the marginal Lp-band detection of 2010), using NACO only. The continuous line represents the combined parallactic motion and the proper motion of the background object in the reference frame of the primary target. The filled dots represent the positions of the candidate companion with error bars at the different epochs while the empty dots (``IF BKGD'') represent the positions of a putative background object at the same epochs. \label{sz82_final_astrometry}}
\end{figure*}

With these proper motions of IM~Lup, we would expect a background source to have moved by $61\pm 9$ mas with respect to IM~Lup between the two observations (NICMOS-2005 and NACO-2008). Such a motion is not detected within our error bars, meaning to first order that the companion is likely co-moving with IM~Lup (Fig.~\ref{sz82_final_astrometry}, left). However, the average SNR on the NICMOS detection, the very slow proper motion and the large astrometric uncertainty mentioned above, do not allow us to firmly and definitely conclude on its bound character. Note that a galactic starcount model for the direction toward IM~Lup \citep{Girardi2005} yields a surface density of stars with $19<Ks<19.5$ of $\simeq 2.2\times 10^5$ per square degree. This makes the chance of a random background source being located within $1\farcs 8$ of IM~Lup $\simeq 17 \%$.

Summer 2011 was the first opportunity to firmly get closure on the bound aspect of this discovered candidate. We then used the epoch 4 NACO observation to redo the common proper motion astrometric analysis. This time, since the analysis is based on a single, well calibrated instrument, our astrometric precision can be trusted down to a conservative $\simeq 10$ mas per coordinate. With a time baseline of 1210 days, the background object should have moved by $68\pm 10$ mas with respect to IM~Lup, which is about the observed variation of separation ($74\pm 20$ mas) in the same direction. We conclude that the candidate companion is likely to be a background object, and is therefore not associated with IM~Lup (Fig.~\ref{sz82_final_astrometry}, right).

\subsection{Probable nature of the point source}\label{subsec:nature}
Based on the combined H, Ks, Lp photometry, we analyzed the SED of the likely background object to verify that it is consistent with a blackbody. For that, we first checked that the 
extinction in the direction of IM~Lup ($\simeq 25^\circ$ from the galactic center bulge) is very small and can indeed be neglected in the near-infrared: $A_H\simeq 0.3$, $A_K\simeq 0.2$, and $A_{Lp}\simeq 0.1$ \citep{Schlegel1998,Schlafly2011}. The SED would be compatible with many possible stellar objects. For instance, the fit to a 3000 K blackbody is satisfying, with the Lp-band point falling only a little more than one sigma above the model. Any blackbody warmer than about 3000 K would actually fit the SED in almost the same way. The reduced $\chi^2$ of the fit is $\simeq 2$ when both the size and temperature of the blackbody are simultaneously fitted. If the background object is actually an M5V star, it would be located far away in our galaxy ($\simeq 4$kpc).

Finally, confirming the stellar nature of the point source, we verified that the H--Ks color derived from our data is not consistent with usual evolutionary models for planetary-mass objects \citep{Baraffe2003, Fortney2008}, which are generally much redder. Note that the Ks--Lp color of $\simeq 1.5-2$ we measured is marginally consistent with both hot start and core accretion models, but the error bars are such that the Lp photometric point is not significant, hence constraining.

\section{Age, distance, and detection limits}\label{subsect:age}

Preliminary age estimates for IM~Lup range from 0.1 Myr to 10 Myr \citep{Hughes1994}. To reduce the uncertainty associated with this large range, we re-estimated the age of IM~Lup as follows. We placed the object on a Hertzsprung-Russell diagram (HRD). The effective temperature was given by the spectral type (conversion from \citet{Luhman2003} for M dwarfs). I and J magnitudes, which are not too much affected by accretion nor disk emission, were converted to bolometric magnitudes based on bolometric corrections and intrinsic colors of \citet{Kenyon1995}.

The bolometric luminosity was then deduced using an estimated distance most probably comprised between 140\,pc \citep{Hughes1994} and 190\,pc \citep{Wichmann1998}\footnote{Recently confirmed by Galli et al.~2012 (in preparation), who measured a kinematic distance of 179 pc for IM~Lup.}, and corrected for extinction using an $Av=0.5$ \citep{Pinte2008} with the law presented in \citet{Draine2003}. We then used evolutionary models from \citet{Baraffe1998} and \citet{Siess2000} to draw isochrones and evolutionary tracks in the HRDs and to interpolate for the observed object. The age estimation was performed independently for I and J photometry, and then folded into error bars. We arrived at the following estimates: for a distance of 140 pc, we get an age range of 0.8-1.75 Myr, while for a distance of 190 pc, we get 0.5-1 Myr. 

\begin{figure}[h]
  \centering
\includegraphics[scale=0.3]{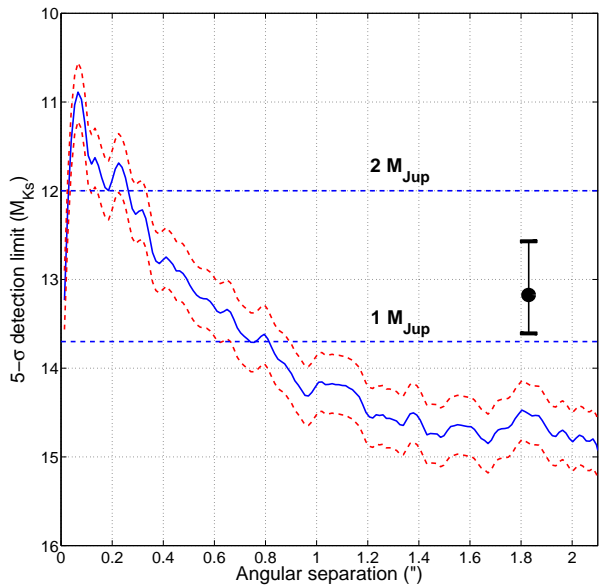}
  \caption{$5\sigma$ detection limit around IM~Lup, derived from the Ks-band 2008 data, assuming an age of $\simeq 1$ Myr and the COND03 model from \citet{Baraffe2003}. The dashed curves show the limits for distances of 140 and 190 pc, respectively. The dashed lines shows the level of contrast for two planet masses: 1, 2 $M_{\rm Jup}$. The black circle and associated error bar shows the point-source Ks band absolute magnitude if associated. \label{sz82_detectionlimit}}
\end{figure}

Our 2008 FQPM data set taken in the Ks band features the best contrast ever achieved around IM~Lup. This data set is therefore suitable to derive detection limits for IM~Lup (Fig.~\ref{sz82_detectionlimit}). We proceeded as follows. For increasing angular separations, we derived the standard deviation in annuli 1 resolution element wide. This profile was then multiplied by 5 to derive the $5\sigma$ detection limit associated with the corresponding data set. 

In Fig.~\ref{sz82_detectionlimit}, we also overplot the level of contrast for two planet masses: 1, 2 $M_{\rm Jup}$, assuming a median age of $\simeq 1$ Myr and considering the COND03 model from \citet{Baraffe2003}. Thanks to the young age of IM~Lup, our (model-dependent) detectability limits are excellent, down to less than one Jupiter mass beyond one arcsecond, and less than two Jupiter masses beyond $0\farcs 2$. Note that the ``core accretion'' model of \citet{Fortney2008} yields much higher masses, reflecting the large uncertainties still plaguing evolutionary models for early ages \citep{Marley2007}.

\section{Conclusion}\label{sec:conclusion}
This paper presented a planet search we conducted with VLT/NACO around the young T~Tauri star IM~Lup between 2008 and 2011, using a pre-discovery image obtained with HST/NICMOS in 2005. IM~Lup is the perfect prototype system for planet search since it has a massive optically thick circumstellar disk, likely at the stability limit. It also features a break in the gas and dust density at about 400 AU, which could indicate the presence of a Jupiter-mass body at the location of the discontinuity. A candidate companion was detected by NACO in 2008, and also seen in the 2005 HST/NICMOS data. 

The candidate companion is located to the North-East of IM~Lup, at a radius of $\simeq 1\farcs 8$, and a position angle (PA) of $\simeq 58^\circ$. Tentatively and naively assuming association, this corresponds to a de-projected physical separation of about 350-480 AU at 140-190 pc. With our redetermined age of about 1 Myr, the mass of the putative off-axis companion using the usual ``hot start'' evolutionary models \citep{Baraffe2003, Fortney2008} would be between $1-2 M_{\rm Jup}$.

However, and unfortunately, the candidate was later on proven to be a background object based on the NACO 2011 observations, a common proper motion analysis and a careful calibration of the NACO plate scale and detector orientation. This cautionary tale taught us the difficulty of planet search around young, distant and obscured stars, where proper motion might not be very well constrained, and where the age and distance determinations are tricky. 

\begin{acknowledgements}
This work was carried out at the European Southern Observatory (ESO) site of Vitacura (Santiago, Chile), and the Jet Propulsion Laboratory (JPL), California Institute of Technology (Caltech), under contract with the National Aeronautics and Space Administration (NASA). OA and JS acknowledge support from the Communaut\'e fran\c{c}aise de Belgique - Actions de recherche concert\'ees - Acad\'emie universitaire Wallonie-Europe. This research has made use of the NASA/IPAC/NExScI Star and Exoplanet Database, which is operated by the JPL, Caltech, under contract with NASA, and NASA's Astrophysics Data System and of the SIMBAD database, operated at CDS (Strasbourg, France). 
\end{acknowledgements}

\bibliographystyle{aa} 
\bibliography{Imlup_dmawet_20120728_final} 
\end{document}